\documentclass[
superscriptaddress,
groupedaddress,
preprintnumbers,
nofootinbib,
amsmath,amssymb,
aps,
prd,
floatfix,
]{revtex4-2}
\usepackage[utf8]{inputenc}
\usepackage{graphicx}
\usepackage{dcolumn}
\usepackage{bm}
\usepackage{hyperref}

\hypersetup{
    colorlinks=true,
    allcolors=blue,
}

\newcommand{\eq}[1]{Eq.~\eqref{#1}}

\usepackage{xcolor}
\newcommand{\lh}[1]{{#1}}

\begin{document}

\title{Dynamic Aspects of Bumblebee Gravity: Post-Newtonian Approach}
\author{Jie Zhu}
 \email{jiezhu@cqu.edu.cn}
 \affiliation{Department of Physics and Chongqing Key Laboratory for Strongly Coupled Physics, Chongqing University, Chongqing 401331, P.R. China}

\author{Hao Li}
  \email{Corresponding author: haolee@cqu.edu.cn}
   \affiliation{Department of Physics and Chongqing Key Laboratory for Strongly Coupled Physics, Chongqing University, Chongqing 401331, P.R. China}

\date{\today}

\begin{abstract}

In this work, we investigate the dynamic aspects of Bumblebee gravity via the parameterized post-Newtonian method.
We find that the PPN framework is self-consistent up to 1.5PN order if and only if $\lambda = -\xi/2$, which corresponds to a direct coupling between the Bumblebee field $B_\mu$ and the Einstein tensor.
The requirement of tachyonic stability restricts the Bumblebee potential to satisfy $V''(0)=0$.
In the specific case where $\lambda = -\xi/2$, the resulting PPN metric yields non-vanishing values for the parameters $\alpha_1$ and $\alpha_2$, as well as a novel PPN potential $U_B$ that exhibits a logarithmic asymptotic growth.
The vanishing of the potential $U_B$ necessitates the additional constraints $\xi = \kappa/2$ or $V^{(3)}(0) = 0$.
These results signify the presence of preferred-frame effects, a direct consequence of the Lorentz symmetry breaking in the model.
In the limit of small $\ell$, we obtain $\alpha_2 \simeq -\ell_2$, which yields a constraint of $|\ell| \lesssim 1.6\times 10^{-9}$ based on pulsar timing observations.

\end{abstract}

\maketitle

\section{Introduction}


The Standard Model (SM) of particle physics and General Relativity (GR) are two fundamental theories describing the natural world: SM addresses particles and quantum interactions, while GR describes classical gravity.
Unifying these theories is crucial for comprehensively understanding nature, leading to various proposed quantum gravity (QG) theories.
Directly testing QG is challenging due to the required Planck scale energies, but potential signals, such as Lorentz symmetry breaking, might be detectable at lower energy scales.

Originating from string theory~\cite{Kostelecky:1988zi}, the Standard-Model Extension (SME)~\cite{Colladay:1996iz, Colladay:1998fq, Colladay:2001wk, Kostelecky:2000mm, Kostelecky:2003fs} offers a comprehensive effective field theory to parameterize these violations. 
A prominent realization of this framework in the gravity sector is the Bumblebee model~\cite{Kostelecky:2003fs, Bumblebee05, Bumblebee08}. 
Here, Lorentz symmetry is spontaneously broken when a vector field $B_\mu$ is driven by a potential to settle into a non-vanishing vacuum expectation value (VEV). By defining a fixed background configuration, this VEV introduces a preferred-frame effect directly into the spacetime geometry. 
This mechanism provides a robust dynamical basis for examining the phenomenological consequences of Lorentz violation in gravitational interactions.

The backreaction of this VEV on the gravitational sector gives rise to a rich phenomenology, including modified black hole solutions~\cite{Casana:2017jkc, Li:2025rjv, Zhu:2025fiy, Liu:2025oho, Zhu:2026vae, Ding:2019mal, Bailey:2025oun, Maluf:2020kgf}, cosmological implications~\cite{Capelo:2015ipa, Maluf:2021lwh, Sarmah:2024xwx, Zhu:2024qcm, Lai:2025nyo}, and gravitational waves~\cite{Lai:2025nyo, Amarilo:2023wpn, Liang:2022hxd}.
The first black hole solution discovered in Bumblebee gravity possesses an asymptotically conical structure~\cite{Casana:2017jkc}. This topological deficit significantly perturbs test-particle geodesics, thereby imposing a stringent upper bound on the Lorentz-violating parameter.
However, recent studies have demonstrated that even in the presence of a non-trivial background field, the Schwarzschild metric remains a consistent solution to Bumblebee gravity, provided that the VEV is timelike~\cite{Zhu:2025fiy, Zhu:2026vae}.
Consequently, the approach of constraining Bumblebee gravity solely through static black hole solutions becomes insufficient. It is therefore imperative to investigate the effects of the background field from a dynamical perspective.

Regarding its dynamical aspects, particularly in the weak-field and slow-motion regime, early investigations in the gravity sector of the SME employed a model-independent approach using general coefficient tensors, and preliminary conclusions for Bumblebee gravity were derived by projecting its specific field configurations onto the broader results of the model-independent framework~\cite{Bailey:2006fd}.
However, such model-independent approaches encounter two primary limitations when applied to specific dynamical models. First, the conservation of the energy-momentum tensor is typically enforced by hand at the linear order. In principle, once the background field dynamics are specified, the energy-momentum conservation should emerge as a natural consequence of the background equations of motion, as dictated by Noether’s theorem. Second, this linear-order enforcement restricts the metric solutions to the 1.5PN order, preventing extensions to higher orders such as 2PN. This limitation is critical because, in the context of particle geodesic motion, the 2PN contributions from $g_{00}$ are physically as significant as $g_{0i}$ at the 1.5PN order.

In this work, we employ a self-consistent approach within the post-Newtonian framework to investigate the weak-field effects of Bumblebee gravity. This is achieved by simultaneously solving the equations of motion for both the background vector field and the metric.
We find that the mathematical self-consistency and physical stability of the PPN formalism provide explicit and definitive constraints on both the model parameters and the potential functions of the vector field in Bumblebee gravity.
By extending the post-Newtonian expansion to the 2PN order, we present a complete parameterized post-Newtonian profile. Remarkably, our high-order analysis uncovers an exotic PPN potential that grows logarithmically with distance, which goes beyond the conventional PPN formulation. Demanding the elimination of this divergent potential consequently imposes severe additional restrictions on the model's parameters.
Furthermore, we observe that the presence of matter causes the background field to deviate from its VEV, marking a distinctive departure from the behavior of vacuum solutions.

The remainder of this paper is organized as follows.
In Sec.~\ref{sec:intro2}, we introduce the Bumblebee gravity action and summarize the PPN expansion scheme.
Section~\ref{sec:cons} derives necessary constraints on the potential $V(X)$ and the coupling parameters, including the conditions $V''(0)=0$ and $\lambda=-\xi/2$.
The full PPN solutions up to 2PN order are presented in Sec.~\ref{sec:PPNSol}, along with the resulting PPN parameters and the new potential $U_B$.
Section~\ref{sec:disscussion} provides a detailed discussion, including the physical implications of $U_B$, the deviation of the vector field from its VEV in the presence of matter, and a comparison with isotropic SME gravity, and Sec.~\ref{sec:summary} summarizes our main findings.

Throughout this paper, Greek indices (e.g., $\mu, \nu$) denote four-dimensional spacetime components, while Latin indices from the beginning of the alphabet (e.g., $a, b$) represent three-dimensional spatial components. The index $0$ is reserved for the temporal component. We adopt the metric signature $(-, +, +, +)$.


\section{Bumblebee Model and PPN Formalism}\label{sec:intro2}

\subsection{The Bumblebee Action and Equations of Motion}

Here, we consider the action of the general bumblebee gravity as 

\begin{equation}
S=\int d^4x\sqrt{-g}\mathcal{L}+S_\mathrm{m},
\end{equation}

\begin{equation}
\begin{aligned}
\mathcal{L}=
&\frac{1}{2\kappa}\left(R+\lambda B_\mu B^\mu R + \xi B^\mu B^\nu R_{\mu\nu}\right)
-\frac{1}{4}B_{\mu\nu}B^{\mu\nu}-V(X).
\end{aligned}
\end{equation}
where $g$ is the determinant of the metric $g_{\mu\nu}$, the constant $\kappa\equiv8\pi G$ with $G$ being the gravitational constant, $S_{\mathrm{m}}$ represents the action for matter fields of no interest in this work, $B_\mu$ is the bumblebee field, and the field strength tensor is $B_{\mu\nu}=\partial_{\mu}B_{\nu}-\partial_{\nu}B_{\mu}$.
In bumblebee theories, the potential $V$ is selected to provide a non-vanishing VEV for $B_\mu$, and could have the following general functional form
\begin{equation*}
    V(X)\equiv V(B^\mu B_\mu + s b^2),\label{potential}
\end{equation*}
where $b$ is a positive real constant, and $s=\pm 1$ or 0 to determine whether the expection value of $B_\mu$ is timelike, spacelike or lightlike. 
In the literature, it is usually assumed that \(V\) has (at least one of) its minimum/maximum at \(0\), thus
\begin{equation*}
    V(0)=0,\ \text{and}\ V'(0)=0.\label{vacuumcondition}
\end{equation*}
The VEV of the bumblebee field is determined when $V(B^\mu B_\mu + s b^2)=0$, implying that
\begin{equation*}
    B^\mu B_\mu+s b^2=0, 
\end{equation*}
The above equation provides a non-null vacuum expectation value
\begin{equation*}
    \langle B^\mu\rangle=b^\mu,
\end{equation*}
where $b_\mu b^\mu + s b^2=0$.
The equation of motions for $g_{\mu\nu}$ is $G_{\mu\nu}=R_{\mu\nu}-\frac{1}{2}Rg_{\mu\nu}=\kappa T_{\mu\nu}$, or
\begin{equation}
R_{\mu\nu}-\kappa(T_{\mu\nu}-\frac{1}{2}g_{\mu\nu}T)=0, \label{eq:met}
\end{equation}
where $T_{\mu\nu}=T_{\mu\nu}^{\mathrm{M}}+T_{\mu\nu}^{B}$,
$T_{\mu\nu}^{\mathrm{M}}$ is the energy-momentum tensor of matter, and 
\begin{widetext}
\begin{equation}
\begin{aligned}
T_{\mu\nu}^{B}=&
-B_{\mu\alpha}B_{\nu}^{\alpha}
-\frac{1}{4}B_{\alpha\beta}B^{\alpha\beta}g_{\mu\nu}
-Vg_{\mu\nu}+2V^{\prime}B_{\mu}B_{\nu}
+\frac{\xi}{\kappa}\left(
\frac{1}{2}B^{\alpha}B^{\beta}R_{\alpha\beta}g_{\mu\nu}
-B_{\mu}B^{\alpha}R_{\alpha\nu}
-B_{\nu}B^{\alpha}R_{\alpha\mu}\right.
\\&
+\frac{1}{2}\nabla_{\alpha}\nabla_{\mu}\left(B^{\alpha}B_{\nu}\right)
+\frac{1}{2}\nabla_{\alpha}\nabla_{\nu}\left(B^{\alpha}B_{\mu}\right)
\left.
-\frac{1}{2}\nabla^{2}\left(B_{\mu}B_{\nu}\right)
-\frac{1}{2}g_{\mu\nu}\nabla_{\alpha}\nabla_{\beta}\left(B^{\alpha}B^{\beta}\right)\right)
\\&
+\frac{\lambda}{\kappa}\left(
-B_\alpha B^\alpha G_{\mu\nu}
-B_\mu B_\nu R
\right.
\left.
+2\nabla_\nu(B_\alpha \nabla_\mu B^\alpha)
-2g_{\mu\nu} \nabla_\alpha(B_\beta \nabla^\alpha B^\beta)\right).
\end{aligned}
\end{equation}
\end{widetext}
For the $B_\mu$ sector, the equation of motion is
\begin{equation}
\nabla^{\mu}B_{\mu\nu}
+\frac{1}{\kappa}(\xi B^\mu R_{\mu \nu}+\lambda B_\nu R)-2V' B_\nu =0.\label{eq:B}
\end{equation}
In this work, we focus on the case where the VEV of the bumblebee field is timelike.

\subsection{Brief Introduction to the PPN Formalism}

The PPN~\cite{Nordtvedt:1968qs, Thorne:1971iat, Will:1971zzb, Will:1971wt, Will:2014kxa, Will_1993, PoissonWill2014} formalism provides a powerful and model-independent framework to bridge the gap between theoretical predictions and weak-field observations. 
Following the setup in the previous section, we now expand the field equations to the second post-Newtonian order within the PPN formalism.
Assuming that the gravitating source matter is constituted by a perfect fluid that obeys the post-Newtonian hydrodynamics, we can write the energy-momentum tensor as
\begin{equation}
T^{\mu\nu}=\begin{pmatrix}\rho+\rho\Pi+p\end{pmatrix}u^{\mu}u^{\nu}+pg^{\mu\nu},\label{eq:emt}
\end{equation}
where $\rho$ is the rest energy density, $\Pi$ is the specific internal energy, $p$ is the pressure, $u^\mu=u^0 (1,v^i)$ is the four-velocity, and its three-velocity is $v^i$.
The energy-momentum tensor~(\ref{eq:emt}) is adequate to derive the post-Newtonian expansion of the gravitational field surrounding a fluid body, such as the Sun, or within a compact binary system~\cite{PoissonWill2014}.
When expanding the energy-momentum tensor in powers of $v/c$, we find that the equation of state and the internal energy are of order two, while the time derivative is of order one; specifically, $p/\rho\sim\Pi\sim(v/c)^{2}$ and $\partial_t\sim(v/c)$~\cite{PoissonWill2014, Will:2018bme}. 
Thus, the energy-momentum tensor~(\ref{eq:emt}) can be expanded in the form
\begin{equation}
\begin{aligned}&T_{00}=\rho\left(1+\Pi+v^{2}-h_{00}^{(2)}\right)+\mathcal{O}(6),\\&T_{0j}=-\rho v_{j}+\mathcal{O}(5),\\&T_{ij}=\rho v_{i}v_{j}+p\delta_{ij}+{\mathcal{O}}(6).\end{aligned}
\end{equation}

With respect to the metric $g_{\mu\nu}$, we expand it around a flat Minkowski background,
\begin{equation}
g_{\mu\nu}=\eta_{\mu\nu}+h_{\mu\nu}=\eta_{\mu\nu}+h_{\mu\nu}^{(1)}+h_{\mu\nu}^{(2)}+h_{\mu\nu}^{(3)}+h_{\mu\nu}^{(4)}+\mathcal{O}(5).
\end{equation}
The generic expansion of the metric components read~\cite{PoissonWill2014}
\begin{equation}
\begin{aligned}
&g_{00}=-1+h_{00}^{(2)}+h_{00}^{(4)}+\mathcal{O}(6),\\
&g_{0i}=h_{0i}^{(3)}+h_{0i}^{(5)}+\mathcal{O}(5),\\
&g_{ij}=\delta_{ij}+h_{ij}^{(2)}+\mathcal{O}(4).
\end{aligned}
\end{equation}
For the vector field $B_\mu$, following the instruction from Will~\cite{Will:2018bme}, we expand $B_\mu$ in a preferred frame, such that $B_\mu^{(0)} = (b, 0, 0, 0)$.
In accordance with the principle of time-reversal invariance, we have
\begin{equation}
\begin{aligned}
B_0&=b+B_0^{(2)}+B_0^{(4)}+\mathcal{O}(6),\\
B_i&=B_i^{(1)}+B_i^{(3)}+\mathcal{O}(5).
\end{aligned}
\end{equation}
In Will’s original formalism, $B_i^{(1)}$ is set to zero; however, we shall initially retain this term here.
To get the post-Newtonian expansion of the field equation, we use the \texttt{xPPN} package developed in Mathematica by Hohmann~\cite{Hohmann:2020muq}.
In the following, we denote the equations of motion by the calligraphic symbol $\mathcal{E}$; specifically, $\mathcal{E}_{\mu\nu}$ represents the metric equations, while $\mathcal{E}_\mu$ denotes the equations of motion for the vector field.

\section{Constraints from PPN Consistency and Physical Instability}\label{sec:cons}

For general parameters and an arbitrary potential, a preliminary PPN analysis reveals inherent inconsistencies and physical instabilities. 
In the following, we demonstrate these issues in detail and subsequently derive initial constraints on the Bumblebee gravity framework.

\subsection{\texorpdfstring{Constraints on the Potential $V(X)$}{Constraints on the Potential V(X)}}

The potential $V(X)$ plays a pivotal role in Bumblebee gravity, as it triggers the spontaneous Lorentz symmetry breaking that endows the vector field with a non-trivial VEV.
In most existing treatments, the potential inducing spontaneous Lorentz symmetry breaking is chosen to resemble the Higgs-type potential, 
{\it i.e.}, the quartic ``Mexican-hat” form $(\phi^2-v^2)^2$ potential. 
Correspondingly, it is taken to be a quadratic function of the invariant quantity $X = B^\mu B_\mu + sb^2$.
That means $V(0)=0$, $V'(0)=0$ and $V''(0)\geq 0$.
However, recent research~\cite{Zhu:2026hxm} suggests a more nuanced perspective: derived from the first principles of SSB, the vacuum state must correspond to the minimum of the Hamiltonian. Owing to the unique constraint structure inherent in vector-tensor theories, the resulting SSB exhibits features that are fundamentally distinct from those observed in the familiar scalar potential models.
The result suggests that for SSB of a vector field in flat spacetime, the extra condition $V''(0) = 0$ and $V^{(3)}(0)\geq 0$ must be satisfied, and the VEV of the Bumblebee field must be either timelike or lightlike.
In the search for vacuum solutions, given the static nature of the metric and the Bumblebee field, the conditions $V(0)=0$ and $V'(0)=0$ ensure that the specific form of the potential $V(X)$ does not affect the structure of the vacuum solutions, as evidenced by the equations of motion.
In a dynamical context, however, where the metric and the vector field evolve under the influence of matter, the functional form of $V(X)$ explicitly enters the solutions. 
We shall proceed to show, via a PN expansion, that $V''(0) = 0$ remains a necessary requirement for the theory's dynamical stability. Even within the context of curved geometries, any deviation from this condition would reintroduce tachyonic pathologies.

We start from the zeroth-order equations.
The non-vanishing components of the equations are
\begin{equation}
\begin{aligned}
\mathcal{E}_{00}^{(0)}=& \kappa \left(V(0)-b^2 V'(0)\right)=0,\\
\mathcal{E}_{ab}^{(0)}=&-\kappa \delta_{ab}\left(V(0)+b^2 V'(0)\right)=0,\\
\mathcal{E}_{0}^{(0)}=&-2 b V'(0)=0.
\end{aligned}
\end{equation}
So we have $V(0)=V'(0)=0$.
The physical implication is straightforward: $V(0)$ effectively acts as a cosmological constant, which must vanish since the PPN formalism is expanded around a Minkowski background. 
Furthermore, $V'(0)$ is associated with the effective mass of the vector field; to allow for a non-trivial field configuration $b$, this mass term is required to be zero.
The only non-vanishing component of the first-order equations is
\begin{equation}
\mathcal{E}_{a}^{(1)}=\Delta B_{a}^{(1)}-\partial_a\partial_b B_{b}^{(1)}=0,\label{eq:0.5PN}
\end{equation}
where summation over repeated indices is implied,
and \(\Delta\equiv \partial_a\partial_a\) is the Laplacian operator.
In the presence of a vector field source, the right-hand side of the first-order equation is identified with the source term. In the absence of the source, $B_{a}^{(1)}=0$ is exactly a solution, consistent with the prescription suggested by Will~\cite{Will:2018bme}.

With the condition and the solution from the zeroth-order and first-order equations, the second-order equations are
\begin{widetext}
\begin{equation}
\begin{aligned}
\mathcal{E}_{00}^{(2)}=&
\left((\lambda+\xi)b^2-1\right)\Delta g_{00}^{(2)}
+\lambda b^2 \left(\partial_a\partial_b g_{ab}^{(2)}-\Delta g_{aa}^{(2)} \right)
+(\xi-2\lambda)b\Delta B_{0}^{(2)}
\\&
+4\kappa b^3 V''(0) B_0^{(2)}
+2\kappa b^4 V''(0) g_{00}^{(2)}
-\kappa \rho=0,
\\
\mathcal{E}_{ab}^{(2)}=&
(\lambda b^2-1)\left(
\partial_a\partial_b g_{cc}^{(2)}+\Delta g_{ab}^{(2)}
-\partial_a\partial_c g_{bc}^{(2)}-\partial_b\partial_c g_{ac}^{(2)}
\right)
+(\lambda b^2+1)\partial_a\partial_b g_{00}^{(2)}
+4\lambda b\; \partial_a\partial_b B_{0}^{(2)}
\\
&+\delta_{ab}\Big(
(\xi+2\lambda)b^2\Delta g_{00}^{(2)}
+\lambda b^2\left(\partial_c\partial_d g_{cd}^{(2)}-\Delta g_{cc}^{(2)}\right)
+(\xi+2\lambda)b \Delta B_0^{(2)}
\\&\quad\quad\quad
+4\kappa b^3 V''(0) B_0^{(2)}
+2\kappa b^4 V''(0) g_{00}^{(2)}
-\kappa \rho
\Big)=0,
\\
\mathcal{E}_{0}^{(2)}=&
(\xi+2\lambda)b\;\Delta g_{00}^{(2)}
+2\lambda b \left( 
\partial_a\partial_b g_{ab}^{(2)}-\Delta g_{aa}^{(2)}
\right)
+2\kappa \Delta B_0^{(2)}
+8\kappa b^2 V''(0) B_0^{(2)}
+4\kappa b^3 V''(0) g_{00}^{(2)}=0.
\end{aligned}\label{eq:1PN}
\end{equation}
Using a similar Fourier transformation method performed in Ref.~\cite{Zhu:2026jkf}, we find that when $V''(0) \neq 0$, both the vector and gravitational sectors exhibit a dual-mode structure, consisting of one massless mode and one massive mode.
The mass of the massive mode is
\begin{equation}
m^2=-\frac{4 \kappa b^2   \left(\lambda b^2  -1\right)  \left((\kappa +2 \lambda ) b^2 -2\right)}{b^2 \left(\lambda b^2  -1\right) (6 \lambda -\xi ) (2 \lambda +\xi )+2 \kappa  \left(\lambda(3 \lambda +\xi ) b^4   -\xi b^2  +1\right)}V''(0).
\end{equation}
\end{widetext}
Considering that $\ell \equiv \xi b^2$ characterizes the strength of Lorentz violation, we can perform an expansion of the above expression in the weak Lorentz-breaking regime ($\ell \ll 1$). The resulting leading-order expression is given by
\begin{equation}
m^2 = -4b^2 V''(0) + \mathcal{O}(\ell^4).
\end{equation}
If the quadratic potential is applied, we have $V''(0)> 0$, which means $m^2<0$.
Notably, the resulting mass term for these modes is negative, which signifies that this massive mode is a tachyon. The presence of such a tachyonic mode across both the gravitational and vector sectors suggests that the Bumblebee gravity, under the quadratic potential, is inherently unstable and potentially unphysical.
To eliminate the tachyonic instability, we must impose the condition $V''(0) = 0$ to ensure the mode remains massless. 


\subsection{\texorpdfstring{Constraints on the Parameters $\lambda$ and $\xi$}{Constraints on the Parameters lambda and xi}}

In the following, we demonstrate that Bumblebee gravity admits a self-consistent PN expansion if and only if the condition $\xi + 2\lambda = 0$ is satisfied. 
Physically, this implies that the non-minimal coupling between the vector field $B_\mu$ and gravity must take the specific form of a direct coupling to the Einstein tensor as $\frac{\xi}{2\kappa} B_\mu B_\nu G^{\mu\nu}$.

This result emerges directly from the procedure of solving the field equations.
In principle, from the 1PN equations~(\ref{eq:1PN}), we can obtain the 1PN solutions for $g_{00}^{(2)}$, $g_{ab}^{(2)}$ and $B_0^{(2)}$.
Nevertheless, for general parameters, a closer inspection reveals that the 1.5PN equations become mathematically inconsistent when the 1PN solutions are applied. 
This internal contradiction points to a potential breakdown of the standard PN expansion for this specific coupling.
Specifically, the existence of the Bianchi identities requires that the divergence of the 1.5PN field equations vanish identically.
For example, for the equations of the Bumblebee field, we have
\begin{equation}
\partial_a \mathcal{E}_{a}^{(3)}=
\partial_0 \left(
\Delta B_0^{(2)}-\frac{\xi b}{2\kappa}\partial_a\partial_b g_{ab}^{(2)} + \frac{\xi b}{2\kappa}\Delta g_{aa}^{(2)}
\right).
\end{equation}
The equation $\partial_a \mathcal{E}_{a}^{(3)}=0$ exhibits a remarkable feature: although it is a third-order equation, all third-order fields cancel out, a property guaranteed by the Bianchi identities. Consequently, this equation effectively becomes a consistency constraint that the second-order solutions must satisfy.
Substituting the 1PN solutions from \eq{eq:1PN} into this constraint, we find that it is satisfied if and only if $\partial_0 \rho\equiv 0$ or $\xi+2\lambda=0$.
This implies that if $\xi + 2\lambda \neq 0$, the PN expansion is restricted to describing scenarios where matter remains static relative to the specific coordinate system where the spatial components of $B_\mu$ vanish. Such a limitation is manifestly unphysical, as it violates the fundamental principles of relativistic gravity. Consequently, we must require $\xi + 2\lambda = 0$ to ensure a consistent PN description.

This condition can also be derived through an alternative approach.
Consider the covariant derivative of the Bumblebee equations of motion $\nabla^\mu \mathcal{E}_{\mu}=0$, with $B_{a}^{(1)}=0$ and $V(0)=V'(0)=V''(0)=0$, we find the first non-vanishing contribution appears at the third order, yielding
\begin{equation}
\left(\nabla^\mu \mathcal{E}_{\mu}\right)^{(3)}=
-\frac{(2\lambda+\xi)b}{2\kappa}\partial_0\left(
\Delta g_{00}^{(2)}+\partial_b\partial_a g_{ab}^{(2)}-\Delta g_{aa}^{(2)}
\right)=0.\label{eq:DE3}
\end{equation}
If $2\lambda+\xi\neq 0$ and the source is dynamical, we have $\Delta g_{00}^{(2)}+\partial_b\partial_a g_{ab}^{(2)}-\Delta g_{aa}^{(2)}=0$.
With $g_{00}^{(2)}=U_1$ and $g_{ab}^{(2)}=\delta_{ab} U_2$, we have $U_2=U_1/2$, which means that the PPN parameter $\gamma=1/2$.
This result is physically untenable, since the PPN parameter $\gamma$ should be a function of the coupling constants $\lambda$ and $\xi$, rather than a fixed constant.
\lh{From a phenomenological perspective, this result is rather unacceptable, since the parameter $\gamma$ has to be extremely close to one.}
So the requirement of PPN consistency necessitates the constraint $\xi + 2\lambda = 0$.
When $\xi + 2\lambda = 0$, \eq{eq:DE3} becomes an identity.
This is inherently due to the fact that under this condition, the non-minimal interaction term in the Bumblebee equations of motion couples the vector field $B_\mu$ directly to the Einstein tensor $G_{\mu\nu}$, and $\nabla^\mu G_{\mu\nu}\equiv 0$ guarantees that $\left(\nabla^\mu \mathcal{E}_{\mu}\right)^{(3)}=0$ is satisfied identically. 

Here, we delve further into the reasons why $\xi + 2\lambda \neq 0$ results in an inconsistent PPN framework.
The origin of this inconsistency stems from the solution of 0.5PN equation~(\ref{eq:0.5PN}).
From a mathematical perspective, the solution to this equation is
\begin{equation}
B_{a}^{(1)} = \partial_a \Phi +B^{\rm T}_a,
\end{equation}
where $\Phi$ is an arbitrary function and $\boldsymbol{B}^{\rm T}$ is a harmonic and divergence-free vector field ($\nabla\cdot\boldsymbol{B}^{\rm T}=0$, $\Delta \boldsymbol{B}^{\rm T}=0$).
If we require $\boldsymbol{B}^{\rm T}$ to vanish at infinity, the unique solution is $\boldsymbol{B}^{\rm T} \equiv 0$.
So the solution is $B_{a}^{(1)} = \partial_a \Phi$.
In a gauge-invariant theory, this solution is a pure gauge; thus, we can choose a gauge such that $B_{a}^{(1)} =0$.
Since the presence of non-minimal coupling breaks the gauge symmetry, we should, in principle, retain $\Phi$. However, within the PPN framework, the solutions at each order must be expressed in terms of PPN potentials, which are determined by the energy-momentum tensor of the matter source. At the 0.5PN order, there are no corresponding potentials available for matching; thus, from the PPN perspective, the solution remains $\Phi=0$.
Should we treat $\Phi$ as an arbitrary function to be determined by higher-order (1PN) solutions and constraints, the resulting system of partial differential equations would become highly non-linear and mathematically intractable. Such a result is fundamentally at odds with the spirit of the PPN framework, which relies on a systematic expansion in terms of matter-sourced potentials. Therefore, we conclude that the PPN approximation for Bumblebee gravity becomes ill-defined whenever the condition $\xi + 2\lambda = 0$ is violated.
Notably, while the 1PN equations admit solutions for arbitrary parameters, the emergence of additional constraints at the 1.5PN order underscores the inherent non-linear nature of gravity.

Alternatively, the fact that 1PN solutions satisfy 1.5PN constraints identically for $\xi + 2\lambda = 0$ suggests that the potentially complex dynamics of $\Phi$ collapse into the trivial solution $\Phi = 0$. Considering the empirical success of the PPN framework in GR and its necessity for precision gravitational tests, it is reasonable to demand that Bumblebee gravity maintains such an expansion. This requirement mandates the condition $\xi + 2\lambda = 0$, thereby identifying the direct coupling between $B_\mu$ and the Einstein tensor as a fundamental structural necessity for the theory's viability.

\section{PPN Solution}\label{sec:PPNSol}

Considering the constraints from the last section, in the following, we shall derive the exact PPN solutions of the Bumblebee gravity with conditions $V(0)=V'(0)=V''(0)=0$ and $\xi+2\lambda=0$ and the 0.5PN solution $B_{a}^{(1)}=0$.

\subsection{1PN Solution}
The 1PN equations are presented as \eq{eq:1PN}.
Substituting $\lambda=-\xi/2$ into \eq{eq:1PN} and using a similar Fourier transformation method performed in Ref.~\cite{Zhu:2026jkf}, we can obtain the solution as
\begin{equation}
\begin{aligned}
g_{00}^{(2)}=&
\frac{\kappa  \left(2+ \xi b^2\right)-4 \xi ^2 b^2 }{2 \pi  \left(2-\xi b^2\right)^2} U,
\\
g_{ab}^{(2)}=&
\frac{\kappa }{2\pi (2-\xi b^2)} \delta_{ab } U,
\\
B_{0}^{(2)}=&-\frac{\xi b}{2\pi (2-\xi b^2)}U,
\end{aligned}
\end{equation}
where
\begin{equation}
U(\mathbf{x},t)\equiv\int\frac{\rho(\mathbf{x}^{\prime},t)}{|\mathbf{x}-\mathbf{x}^{\prime}|}d^3x^{\prime}.
\end{equation}
This 1PN solution provides the effective gravitational coupling constant $G_{\rm eff}$ and the PPN parameter $\gamma$ as
\begin{equation}
\begin{aligned}
G_{\rm eff} = & \frac{\kappa  \left(2+ \xi b^2\right)-4 \xi ^2 b^2 }{4 \pi  \left(2-\xi b^2\right)^2},
\\
\gamma = & \frac{\kappa (2-\xi b^2)}{\kappa  \left(2+ \xi b^2\right)-4 \xi ^2 b^2}.
\end{aligned}
\end{equation}
In the limit $\ell\equiv\xi b^2\ll 1$, they can be approximated as
\begin{equation}
\begin{aligned}
G_{\rm eff} \simeq & G+\frac{3\kappa-4\xi}{16\pi}\ell,
\\
\gamma \simeq & 1 - \frac{\kappa-2\xi}{\kappa}\ell.
\end{aligned}
\end{equation}

Two special cases are discussed here.
The first one is $\xi=\kappa/2$.
In this case, the vacuum solution admits an extra degree of freedom in the Bumblebee field.
When $\xi=\kappa/2$, $G_{\rm eff}$ and $\gamma$ become
\begin{equation}
G_{\rm eff} = \frac{G}{1-\ell /2}, \quad \gamma=1.
\end{equation}
In this case, the PPN parameter $\gamma$ becomes identical to that of GR.
Another case is $(\lambda+\xi)b^2=\xi b^2/2\equiv1$.
In this case, the vacuum solution's entropy vanishes identically, which points to an unphysical situation.
Directly from \eq{eq:1PN}, in this case we have
\begin{equation}
\mathcal{E}_{00}^{(2)}+\delta_{ab}\mathcal{E}_{ab}^{(2)}=-4\kappa\rho =0.
\end{equation}
This suggests that the PPN framework is only consistent in a vacuum for this particular choice of parameter. Specifically, the condition $(\lambda + \xi)b^2 = 1$ restricts the theory to vacuum solutions only, effectively precluding the description of matter sources.
In other words, the requirement for a consistent description of matter sources naturally rules out the parameter range that yields zero black hole entropy~\cite{Zhu:2026vae}, ensuring the theory remains physically viable.

\subsection{1.5PN Solution}

With the solution from 1PN, the 1.5PN equations are
\begin{widetext}
\begin{equation}
\begin{aligned}
\mathcal{E}_{0a}^{(3)}=&
(2-\xi b^2)\left(
\partial_a\partial_b g_{0b}^{(3)}
-\Delta g_{0a}^{(3)}
\right)
-2\xi b\left(
\partial_a\partial_b B_{b}^{(3)}-\Delta B_a^{(3)}
\right)
-\frac{\kappa}{4}\partial_0\partial_a U
+4\kappa \rho v_a=0
,\\
\mathcal{E}_{a}^{(3)}=&
\xi b \left(
\partial_a\partial_b g_{0b}^{(3)}-\Delta g_{0a}^{(3)}
\right)
+2\kappa\left(
\partial_a\partial_b B_{b}^{(3)}-\Delta B_a^{(3)}
\right)=0.
\end{aligned}
\end{equation}
\end{widetext}
From the above two equations, we can solve $\partial_a\partial_b g_{0b}^{(3)}-\Delta g_{0a}^{(3)}$ and $\partial_a\partial_b B_{b}^{(3)}-\Delta B_a^{(3)}$, and the problem becomes analogous to the case in GR.
A more straightforward approach is to solve the equations using the method of undetermined coefficients.
By adopting the ansatz for the solution as
\begin{equation}
\begin{aligned}
g_{0a}^{(3)}=& c_1 V_{a} +c_2 W_{a},\\
B_a^{(3)}=& c_3 V_{a} +c_4 W_{a},
\end{aligned}
\end{equation} 
and substituting it into the equations, we can solve the coefficients as
\begin{equation}
\begin{aligned}
c_1=& m_0 -\frac{\kappa^2}{2\pi \left(2\kappa - \xi b^2 (\kappa-\xi)\right)},\\
c_2=& -m_0 -\frac{\kappa^2}{2\pi \left(2\kappa - \xi b^2 (\kappa-\xi)\right)},\\
c_3=& n_0 + \frac{\kappa\xi b}{4\pi \left(2\kappa - \xi b^2 (\kappa-\xi)\right)},\\
c_4=& -n_0 + \frac{\kappa\xi b}{4\pi \left(2\kappa - \xi b^2 (\kappa-\xi)\right)},\\
\end{aligned}
\end{equation}
where $m_0$ and $n_0$ are gauge parameters, which are typically determined by matching the 2PN solutions with the standard PPN expansion.

\subsection{2PN Solutiuons}

The 2PN equations are shown in Appendix~\ref{sec:2PNeq}.
In the standard PPN formalism, the analysis of 2PN corrections is restricted to $g_{00}$ and the Bumblebee field.
The 2PN equations differ from those in GR or scalar-tensor theories in two key respects.
Firstly, unlike in GR or scalar-tensor theories where $g_{00}^{(4)}$ is governed by a standalone equation $\mathcal{E}_{00}^{(4)}=0$, the Bumblebee gravity framework exhibits a mixing between $g_{00}^{(4)}$, $g_{ab}^{(4)}$, and $B_{0}^{(4)}$. Consequently, one must solve these equations jointly rather than treating them in isolation.
Secondly, the emergence of a novel source term $U^2$ necessitates the introduction of an additional potential. This potential goes beyond the standard PPN formalism and is required to fully characterize the 2PN corrections.

To address the first problem, we observe that the 2PN system can be recast as a set of linear equations in terms of $\Delta g_{00}^{(4)}$, $\Delta B_{0}^{(4)}$, and the combination $\Delta g_{aa}^{(4)}-\partial_a\partial_b g_{ab}^{(4)}$. This allows us to solve for these collective variables first, thereby simplifying the derivation of the 2PN solutions (see Appendix~\ref{sec:2PNeq}).
For the second problem, we need to introduce a new potential $\Phi_B$ as
\begin{equation}
U_B(\mathbf{x},t)\equiv\int\frac{U(\mathbf{x}^{\prime},t)^2}{|\mathbf{x}-\mathbf{x}^{\prime}|}d^3x^{\prime},\label{eq:UB}
\end{equation}
and it satisfies
\begin{equation}
\Delta U_B=-4\pi U^2.
\end{equation}

To solve the 2PN results, we use the method of undetermined coefficients.
Let the solution be
\begin{equation}
\begin{aligned}
g_{00}^{(4)}=& a_1 \Phi_1 +a_2 \Phi_2 +a_3 \Phi_3 +a_4 \Phi_4 +a_5 U^2 +a_6 U_B,\\
B_{0}^{(4)}=& b_1 \Phi_1 +b_2 \Phi_2 +b_3 \Phi_3 +b_4 \Phi_4 +b_5 U^2 +b_6 U_B,
\end{aligned}
\end{equation}
where $\Phi_1$, $\Phi_2$, $\Phi_3$, and $\Phi_4$ are PPN potentials shown in Appendix~\ref{sec:SPPN}.
Substitution into the governing equations yields the following coefficients
\begin{widetext}
\begin{equation}
\begin{aligned}
&a_1=-\frac{2 (\xi  \ell -\kappa )}{\pi  (\ell -2)^2},\quad
a_2=\frac{\kappa  \xi  \ell  \left(5 \ell ^2-2 \ell +16\right)-4 \xi ^2 \ell  \left(\ell ^2+\ell +2\right)-\left(\kappa ^2 \left(\ell ^3+\ell ^2+4\right)\right)}{2 \pi ^2 (\ell -2)^5},\\
&a_3=\frac{\kappa  (\ell +2)-4 \xi  \ell }{2 \pi  (\ell -2)^2},\quad
a_4=-\frac{3 \kappa }{2 \pi  (\ell -2)},\\
&a_5=\frac{4 \kappa  \xi  \ell  \left(2 \ell ^2+5 \ell -2\right)+8 \xi ^2 \ell  \left(-2 \ell ^2+\ell -2\right)-\left(\kappa ^2 \left(\ell ^3+6 \ell ^2+4 \ell -8\right)\right)}{8 \pi ^2 (\ell -2)^5},\quad
a_6=\frac{ b^6 (\ell +2)^3 (\kappa -2 \xi )^3V^{(3)}(0)}{8 \pi ^3 (\ell -2)^6},
\end{aligned}
\end{equation}
\begin{equation}
\begin{aligned}
&b_1= \frac{\xi b }{2 \pi  (\ell -2)},\quad
b_2= \frac{\xi b \left(\xi  \left(\ell ^2+\ell +2\right)-\kappa  \left(\ell ^2-2 \ell +4\right)\right)}{2 \pi ^2 (\ell -2)^4},\\
&b_3= \frac{ \xi b }{2 \pi  (\ell -2)}, \quad
b_4= 0,\\
&b_5= \frac{\xi b   (\xi  \ell -2 \kappa )}{8 \pi ^2 (\ell -2)^3},\quad
b_6= \frac{b^5 (\ell +2)^3 (\kappa -2 \xi )^2 V^{(3)}(0) }{16 \pi ^3 (\ell -2)^5},
\end{aligned}
\end{equation}
\end{widetext}
where $\ell\equiv \xi b^2$.
The 1.5PN parameters $m_0$ and $n_0$ are also determined as
\begin{equation}
m_0=\frac{3 \kappa }{8 \pi  (\ell -2)},\quad
n_0=0.
\end{equation}

\subsection{PPN Parameters} \label{sec:PPNP}
Since the newly introduced potential $U_B$, the standard PPN formalism should be modified as
\begin{equation}
\begin{aligned}
g_{00}^{(2)}=& 2G U,
\\
g_{ab}^{(2)}=&2\gamma GU \delta_{ab},
\\
g_{0a}^{(3)}=& -\frac{1}{2}(3+4\gamma+\alpha_1-\alpha_2+\zeta_1-2\xi)GV_a-\frac{1}{2}(1+\alpha_2-\zeta_1+2\xi)GW_a,
\\
g_{00}^{(4)}=&
-2\beta G^2U^2+(2+2\gamma+\alpha_3+\zeta_1-2\xi)G\Phi_1+2(1+3\gamma-2\beta+\zeta_2+\xi)G^2\Phi_2
\\&
+2(1+\zeta_3)G\Phi_3+2(3\gamma+3\zeta_4-2\xi)G\Phi_4-2\xi G^2\Phi_W-(\zeta_1-2\xi)G\mathcal{A}
+\beta_B U_B,
\end{aligned}
\end{equation}
where $(\beta,\gamma,\xi,\zeta_1,\zeta_2,\zeta_3,\zeta_4,\alpha_1,\alpha_2,\alpha_3)$ are 10 PPN parameters in the standard PPN formalism, and $\beta_B$ is a new parameter introduced to describe the contribution of the potential $U_B$.

Comparing the PN solutions to the Bumblebee gravity with the standard PPN formalism, we can obtain the PPN parameters as
\begin{equation}
\begin{aligned}
G_{\rm eff}=& \frac{\kappa  (\ell +2)-4 \xi  \ell }{4 \pi  ( \ell-2)^2},
\\
\beta=& \frac{-4 \kappa  \xi  \ell  \left(2 \ell ^2+5 \ell -2\right)+8 \xi ^2 \ell  \left(2 \ell ^2-\ell +2\right)+\kappa ^2 \left(\ell ^3+6 \ell ^2+4 \ell -8\right)}{(\ell -2) (\kappa  (\ell +2)-4 \xi  \ell )^2},
\\
\gamma =& -\frac{\kappa  (\ell -2)}{\kappa  (\ell +2)-4 \xi  \ell },
\\
\alpha_1=& -\frac{8 \ell  \left(\kappa ^2 (\ell -2)-2 \kappa  \xi  (\ell -1)+2 \xi ^2 \ell \right)}{\kappa ^2 \left(\ell ^2-4\right)+4 \xi ^2 \ell ^2+\kappa  \xi  (6-5 \ell ) \ell },
\\
\alpha_2=& \frac{2 \ell  \left(-\left(\kappa ^2 (\ell -2)\right)+\kappa  \xi  \ell -2 \xi ^2 \ell \right)}{\kappa ^2 \left(\ell ^2-4\right)+4 \xi ^2 \ell ^2+\kappa  \xi  (6-5 \ell ) \ell },
\\
\beta_B =& \frac{ b^6 (\ell +2)^3 (\kappa -2 \xi )^3V^{(3)}(0)}{8 \pi ^3 (\ell -2)^6},
\\
\xi=&\alpha_3 = \zeta_1= \zeta_2=\zeta_3=\zeta_4=0.
\end{aligned}
\end{equation}
The non-vanishing nature of $\alpha_1$ and $\alpha_2$ is to be expected, given that Bumblebee gravity intrinsically entails Lorentz symmetry breaking.
In the limit $\ell\equiv\xi b^2\ll 1$, the PPN parameters can be approximated as
\begin{equation}
\beta-1 \simeq -\frac{(\kappa -2 \xi ) (\kappa -\xi )}{\kappa ^2}\ell ,
\quad
\gamma-1\simeq -\frac{\kappa - 2 \xi}{\kappa }\ell,
\quad
\alpha_1 \simeq -\frac{4(\kappa-\xi)}{\kappa}\ell,
\quad
\alpha_2 \simeq -\ell.\label{eq:small_limit}
\end{equation}
One special case is $\xi=\kappa/2$, in which the PPN parameters reduce to
\begin{equation}
\beta=\gamma=1, \quad
\alpha_1 = -\frac{2\ell}{1-\ell /2}, \quad
\alpha_2 = -\frac{\ell}{1-\ell /2}, \quad \beta_B=0.
\end{equation}
In this case, the contribution of $U_B$ vanishes.
Even though the PPN parameters $\beta$ and $\gamma$ coincide with those of GR, the preferred-frame parameters $\alpha_1$ and $\alpha_2$ remain non-zero.

While the PPN parameters $\beta$ and $\gamma$ in Bumblebee gravity can be tuned to align with GR, the preferred-frame parameters $\alpha_1$ and $\alpha_2$ generally remain non-vanishing. These non-zero coefficients signify a departure from local Lorentz invariance, thereby providing a robust observational channel to constrain the Bumblebee coupling constants through high-precision solar system experiments and pulsar timing observations.
Currently, the most stringent constraints on these two parameters are provided by solar system experiments and pulsar timing observations: the former yields limits at the level of $|\alpha_1| \lesssim \times 10^{-4}$~\cite{Muller:1996up} and $|\alpha_2| \lesssim 2.4\times 10^{-7}$~\cite{1987ApJ...320..871N}, while the latter pushes these bounds to the scales of $|\alpha_1| \lesssim 4\times 10^{-5}$~\cite{Shao:2012eg} and $|\alpha_2| \lesssim 1.6\times 10^{-9}$~\cite{Shao:2013wga}.
Since the constraints on $\alpha_2$ are significantly more stringent than those on $\alpha_1$, and given that $\alpha_2 \simeq -\ell$ in the limit $\ell \ll 1$, we can utilize these PPN parameters obtain the constraint
\begin{equation}
    |\ell|\lesssim 1.6\times 10^{-9}.
\end{equation}

\section{Discussions}\label{sec:disscussion}

\subsection{Impact of the New Potential}

To provide a complete description of the 2PN expansion in Bumblebee gravity, a novel potential $U_B$ was introduced in the preceding section as \eq{eq:UB}. In this subsection, we delve into a detailed discussion regarding the physical implications and the specific contributions of this newly defined potential.

We consider a source consisting of a homogeneous sphere with density $\rho_0$, radius $R$, and mass $M$, and subsequently evaluate the contribution of this source to the potential $U_B$.
The Newton potential $U(\mathbf{x})$ can be obtained by the convolution $\frac{1}{r}*\rho(r)$.
In three-dimensional Euclidean space, the convolution of $f(r)$ and $g(r)$ can be expressed as 
\begin{equation}
(f*g)(r)=\frac{2\pi}{r}\int_0^\infty dr^{\prime}r^{\prime}g(r^{\prime})\int_{|r-r^{\prime}|}^{r+r^{\prime}}ds\left.s\right.f(s).\label{eq:int}
\end{equation}
Let $f$ be the convolution kernel $1/r$, we obtain
\begin{equation}
U(r)=\frac{1}{r}*\rho(r)=\begin{cases}
 \frac{M}{r} & r\geq R, \\
 \frac{M \left(3 R^2-r^2\right)}{2 R^3} & 0<r< R. 
\end{cases}
\end{equation}
However, when employing the convolution formula to evaluate $U_B$, we find that the integral fails to converge. Consequently, we introduce a cutoff $L$ as the upper limit of integration. The resulting expression is given by
\begin{equation}
U_B(r)=\frac{1}{r}*U(r)^2 =
\begin{cases}
4\pi M^2 \Big(1-\frac{18 R}{35 r}+\log \left(\frac{L}{r}\right)\Big)
& r\geq R, \\
4\pi M^2 \Big(\frac{19}{24}-\frac{3 r^2}{8
   R^2}+\frac{3 r^4}{40 R^4}-\frac{r^6}{168 R^6}+\log \left(\frac{L}{R}\right)\Big)
& 0<r< R. 
\end{cases}
\end{equation}
Subsequently, we can subtract the $L$-dependent divergent terms, leading to the final finite expression
\begin{equation}
\bar{U}_B(r)=
\begin{cases}
4\pi M^2 \Big(1-\frac{18 R}{35 r}-\log \left(\frac{r}{R}\right)\Big)
& r\geq R, \\
4\pi M^2 \Big(\frac{19}{24}-\frac{3 r^2}{8
   R^2}+\frac{3 r^4}{40 R^4}-\frac{r^6}{168 R^6}\Big)
& 0<r< R. 
\end{cases}
\end{equation}
It is easy to check that the finite part $\bar{U}_B$ satisfies the relation $\Delta \bar{U}_B=-4\pi U^2$.

It can be observed that in the exterior region ($r > R$), the leading-order contribution to $U_B$ exhibits a logarithmic dependence on $r$.
This implies that a non-vanishing $\beta_B$ would introduce a logarithmic correction to the 2PN metric, a feature that departs significantly from GR. Consequently, to ensure consistency with the standard GR limit, the condition $\beta_B = 0$ is found to be necessary.
In other words, either the parameters must satisfy the condition $\xi = \kappa/2$, or the potential function must fulfill the requirement $V^{(3)}(0) = 0$.

\subsection{Bumblebee away from the VEV}

In the vacuum solutions of Bumblebee gravity, valid solutions exist under the condition where the vector field is strictly constrained to its VEV~\cite{Casana:2017jkc, Li:2025rjv, Zhu:2025fiy, Liu:2025oho, Zhu:2026vae, Ding:2019mal}.
Furthermore, vacuum solutions in which the vector field deviates from its VEV have also been identified~\cite{Bailey:2025oun}.
However, the presence of matter invariably causes the vector field to deviate from its VEV.
Within the PPN framework, this deviation is given by
\begin{equation}
B_\mu B^\mu +b^2 =
-\Big(2b B_0^{(2)} +b^2 g_{00}^{(2)}\Big)
-\Big(
\left(B_0^{(2)}\right)^2 
+ 2b \left(B_0^{(4)}+B_0^{(2)} g_{00}^{(2)}\right)
\Big)+\mathcal{O}(5).
\end{equation}
Specifically, utilizing the metric and vector field solutions obtained in the preceding section, we can evaluate the explicit deviation of the vector field from its VEV.
When $\xi \neq \kappa/2$, the deviation primarily occurs at the 1PN order, and the result is given by
\begin{equation}
B_\mu B^\mu +b^2 =-\frac{b^2(\kappa-2\xi) (2+\ell)}{2\pi (2-\ell)^2}U+\mathcal{O}(4).
\end{equation}
In this case, the leading-order deviation is proportional to the Newtonian potential.
In the case where $\xi = \kappa/2$, the deviation occurs at the 2PN order, and its explicit form is given by
\begin{equation}
B_\mu B^\mu +b^2 =
-\frac{\ell}{\pi(2-\ell)}(\Phi_1+3\Phi_4)-\frac{\kappa\ell}{4\pi^2(2-\ell)^3}(U^2-2\Phi_2)
+\mathcal{O}(6).
\end{equation}
For a static point-particle source, the deviation is proportional to the square of the Newtonian potential.

Here, we identify a fundamental discrepancy between Bumblebee gravity and GR: for a static point-particle source, the PPN expansion in GR consistently recovers the Schwarzschild vacuum solution. In contrast, Bumblebee gravity fails to do so. This stems from the fact that the presence of matter inevitably drives the vector field away from its VEV, whereas the standard vacuum solutions in Bumblebee gravity can reside exactly at the VEV.

\subsection{Compare with Isotropic SME Gravity}\label{sec:SME}

The SME framework contains the Lagrangian densities for both the Standard Model and GR, along with all scalar terms involving operators for Lorentz violation and CPT violation, offering a general parametrization of Lorentz and CPT violation.
The action of the gravity sector in the SME framework is given by~\cite{Bailey:2006fd}
\begin{equation}
S=\int d^4x\sqrt{-g}\frac{R}{2\kappa}+S_{\mathrm{LV}}+S^{\prime},
\end{equation}
where $S_{\mathrm{LV}}$ contains the leading Lorentz-violating couplings as follows
\begin{equation}
S_{\mathrm{LV}}=\frac{1}{2\kappa}\int d^4x\sqrt{-g}(-uR+s^{\mu\nu}R_{\mu\nu}^\mathrm{T}+t^{\kappa\lambda\mu\nu}C_{\kappa\lambda\mu\nu}),
\end{equation}
where $u, s^{\mu\nu}$, and $t^{\kappa\lambda\mu\nu}$ are the fields inducing Lorentz violation, $R_{\mu\nu}^{\mathrm{T}}$ is the trace-free Ricci tensor, and $C_{\kappa\lambda\mu\nu}$ is the Weyl tensor. 
The tensor field $s^{\mu\nu}$ is taken to be symmetric and traceless. The tensor field $t^{\kappa\lambda\mu\nu}$ inherits the symmetries of the Riemann tensor, and its trace and partial traces vanish. The action $S^{\prime}$ includes dynamics not only for conventional matter but also for the Lorentz-violating fields $u,s^{\mu\nu}$ and $t^{\kappa\lambda\mu\nu}$.

The PPN metric for the SME model is obtained in Ref.~\cite{Bailey:2006fd}.
The coefficient tensor fields $u$, $s^{\mu\nu}$, and $t^{\kappa\lambda\mu\nu}$ contribute an additional energy-momentum tensor. Owing to the Bianchi identities, the existence of solutions to the equations of motion necessitates the conservation of this energy-momentum tensor.
Under the premise that the dynamics of the coefficient tensor fields remain unknown, Ref.~\cite{Bailey:2006fd} adopted additional assumptions by considering linear perturbations around the tensor backgrounds and requiring the energy-momentum corrections to be conserved at the linear order. Based on these assumptions, the metric was subsequently derived.
While this approach provides a preliminary description of the PPN metric in Lorentz-violating gravity, it suffers from two major limitations. First, the enforced energy-momentum conservation holds only at the linear order; consequently, the PPN expansion of the metric can be extended to the 1.5PN level at most, as demonstrated in Ref.~\cite{Bailey:2006fd}. But when considering the geodesic motion of particles, the 1.5PN contributions are as significant as the 2PN contributions to $g_{00}$.
Second, in scenarios where the dynamics of the background fields are explicitly known, the field equations of the background itself ensure that the energy-momentum corrections are automatically conserved. From a consistency standpoint, the correct approach should involve solving the equations for both the background and gravitational fields simultaneously.

Here we compare our results with the PPN parameters for isotropic SME obtained in Ref.~\cite{Bailey:2006fd}.
The interesting result in Ref.~\cite{Bailey:2006fd} is that the PPN parameters $\alpha_1$, $\alpha_2$ and $\gamma$ satisfy the relation
\begin{equation}
\alpha_1=4\alpha_2, \quad \gamma = 1+\alpha_2.
\end{equation}
However, in the small $\ell$ limit, our result~(\ref{eq:small_limit}) suggests that
\begin{equation}
\alpha_1 \simeq 4(1-\frac{\xi}{\kappa})\alpha_2,
\quad
\gamma \simeq 1+ (1-\frac{2\xi}{\kappa})\alpha_2.
\end{equation}
A significant discrepancy exists between these two results, which vanishes only when $\xi=0$. However, in this limit, the effects of Lorentz violation are no longer present in the Bumblebee gravity.
Furthermore, our calculations demonstrate that when $\xi = \kappa/2$, the PPN parameters $\beta$ and $\gamma$ can be strictly unity, with deviations from GR appearing only in $\alpha_1$ and $\alpha_2$. This stands in contrast to the results of isotropic SME, which suggest that if $\alpha_1$ and $\alpha_2$ are non-zero, $\gamma$ must necessarily deviate from unity.

\section{Summary}\label{sec:summary}

In this work, we present a systematic investigation into the dynamical properties of Bumblebee gravity within the PPN framework. 
Our analysis reveals that the self-consistency of the PPN formalism imposes stringent constraints on the theory's parameter space. 
Specifically, mathematical consistency at the 1.5PN order is maintained if and only if the coupling constants satisfy $\lambda = -\xi/2$. 
Physically, this condition corresponds to a direct coupling between the vector field $B_{\mu}$ and the Einstein tensor.
Furthermore, to eliminate potential tachyonic instabilities in both the gravitational and vector sectors, the Bumblebee potential must satisfy the constraint $V''(0)=0$.
This conclusion aligns with recent studies~\cite{Zhu:2026hxm}, in which a Hamiltonian analysis reveals that the SSB of the vector field requires the potential function to be at least cubic.

Under these consistency requirements, we obtain the full PPN metric. We find that the potential $V(X)$ manifests as an additional PPN-like term $U_B$ at the 2PN level. 
Unlike standard potentials that vanish at infinity, $U_B$ grows logarithmically with distance $r$, rather than the conventional power-law decay.
Such a potential induces a radical departure from the GR-like asymptotic structure of the metric. 
Consequently, to suppress this divergent $U_B$ term and maintain asymptotic flatness, the theory is restricted to a specific parameter subspace where either the coupling satisfies $\xi = \kappa/2$ or the potential is constrained by $V^{(3)}(0) = 0$.

The PPN parameters are obtained in Sec.~\ref{sec:PPNP}.
The results indicate that the spontaneous Lorentz violation in Bumblebee gravity leads to significant preferred-frame effects, manifested by non-zero PPN parameters $\alpha_1$ and $\alpha_2$. In the limit of weak Lorentz violation, we derive the relation $\alpha_2 \simeq -\ell$, where $\ell=\xi b^2$.
By incorporating observational data from pulsar timing observations, we establish a stringent constraint on the Lorentz-violating parameter as $|\ell|\lesssim 1.6\times 10^{-9}$.
Furthermore, we find that when the condition $\xi = \kappa/2$ is satisfied, the PPN parameters $\beta$ and $\gamma$ become unity strictly. In this specific parameter regime, the theory's departure from GR is uniquely characterized by non-zero values of $\alpha_1$ and $\alpha_2$.

Additionally, we find that our PPN results exhibit notable discrepancies with those obtained from the isotropic SME~\cite{Bailey:2006fd}. The underlying reasons for these differences are discussed in detail in Sec.~\ref{sec:SME}, primarily arising from the treatment of the energy-momentum tensor conservation. In the isotropic SME, the conservation of the energy-momentum tensor is enforced by hand at the linear order. In contrast, our approach ensures that the energy-momentum tensor is automatically conserved by simultaneously solving the equations of motion for the vector field.
Our approach ensures the energy-momentum conservation to all orders, thereby enabling the extension of the PPN formalism to the 2PN order. This high-order consistency is crucial, as 2PN effects are as significant as those at the 1.5PN order when describing the geodesic motion of particles.

Furthermore, we discuss the impact of matter on Bumblebee gravity, drawing a clear distinction between the solutions in the presence of material sources and those in a vacuum.
The first distinction lies in the fact that the presence of matter excludes the parameter space satisfying $(\lambda+\xi)b^2 = \xi b^2/2 \equiv 1$, under which the equations of motion at the 1PN order admit no solution. Notably, this specific parameter regime is unphysical, as it corresponds to a black hole solution with vanishing entropy~\cite{Zhu:2026vae}.
The second distinction is that the presence of matter causes the vector field to deviate from its VEV, whereas in the vacuum case, the field can reside strictly at the VEV.

In summary, our analysis establishes the necessary self-consistency conditions and PPN signatures for Bumblebee gravity. These insights are not only pivotal for evaluating the theory's viability but also serve as a cornerstone for future high-precision gravity tests, such as those involving pulsar timing and gravitational-wave observations.

\section*{Acknowledgements}

We acknowledge the use of the \texttt{xPPN} package \url{https://github.com/xenos1984/xPPN}~\cite{Hohmann:2020muq} in the theoretical calculations. This work was supported in part by the National Natural Science Foundation of China under Grant No.~12547101. HL was also supported by the start-up fund of Chongqing University under No.~0233005203009, and JZ was supported by the start-up fund of Chongqing University under No.~0233005203006.

\appendix

\section{PPN Potentials}\label{sec:SPPN}
In this Appendix, we present the explicit expressions for the PPN potentials used to parameterize the metric.
These potentials are given as follows~\cite{Will_1993}:
\begin{equation}
\chi(t,\vec{x})=-\int\mathrm{d}^3x^{\prime}\rho(t,\vec{x}^{\prime})|\vec{x}-\vec{x}^{\prime}|,
\end{equation}
\begin{equation}
U(t,\vec{x})=\int\mathrm{d}^3x^{\prime}\frac{\rho(t,\vec{x}^{\prime})}{|\vec{x}-\vec{x}^{\prime}|},
\end{equation}
\begin{equation}
U_{ab}(t,\vec{x})=\int\mathrm{d}^3x^{\prime}\frac{\rho(t,\vec{x}^{\prime})}{|\vec{x}-\vec{x}^{\prime}|^3}(x_a-x_a^{\prime})(x_b-x_b^{\prime}),
\end{equation}
\begin{equation}
V_a(t,\vec{x})=\int\mathrm{d}^3x^{\prime}\frac{\rho(t,\vec{x}^{\prime})v_a(t,\vec{x}^{\prime})}{|\vec{x}-\vec{x}^{\prime}|},
\end{equation}
\begin{equation}
W_a(t,\vec{x})=\int\mathrm{d}^3x^{\prime}\frac{\rho(t,\vec{x}^{\prime})v_b(t,\vec{x}^{\prime})(x_a-x_a^{\prime})(x_b-x_b^{\prime})}{|\vec{x}-\vec{x}^{\prime}|^3},
\end{equation}
\begin{equation}
\Phi_1(t,\vec{x})=\int\mathrm{d}^3x^{\prime}\frac{\rho(t,\vec{x}^{\prime})v(t,\vec{x}^{\prime})^2}{|\vec{x}-\vec{x}^{\prime}|},
\end{equation}
\begin{equation}
\Phi_2(t,\vec{x})=\int\mathrm{d}^3x^{\prime}\frac{\rho(t,\vec{x}^{\prime})U(t,\vec{x}^{\prime})}{|\vec{x}-\vec{x}^{\prime}|},
\end{equation}
\begin{equation}
\Phi_3(t,\vec{x})=\int\mathrm{d}^3x^{\prime}\frac{\rho(t,\vec{x}^{\prime})\Pi(t,\vec{x}^{\prime})}{|\vec{x}-\vec{x}^{\prime}|},
\end{equation}
\begin{equation}
\Phi_4(t,\vec{x})=\int\mathrm{d}^3x^{\prime}\frac{p(t,\vec{x}^{\prime})}{|\vec{x}-\vec{x}^{\prime}|},
\end{equation}
\begin{equation}
\mathcal{A}(t,\vec{x})=\int\mathrm{d}^3x^{\prime}\frac{\rho(t,\vec{x}^{\prime})\left[v_a(t,\vec{x}^{\prime})(x_a-x_a^{\prime})\right]^2}{|\vec{x}-\vec{x}^{\prime}|^3},
\end{equation}
\begin{equation}
\mathcal{B}(t,\vec{x})=\int\mathrm{d}^3x^{\prime}\frac{\rho(t,\vec{x}^{\prime})}{|\vec{x}-\vec{x}^{\prime}|}(x_a-x_a^{\prime})\frac{\mathrm{d}v_a(t,\vec{x}^{\prime})}{\mathrm{d}t},
\end{equation}
\begin{equation}
\Phi_W(t,\vec{x})=\int\mathrm{d}^3x^{\prime}\int\mathrm{d}^3x^{\prime\prime}\rho(t,\vec{x}^{\prime})\rho(t,\vec{x}^{\prime\prime})\frac{x_a-x_a^{\prime}}{|\vec{x}-\vec{x}^{\prime}|^3}\left(\frac{x_a^{\prime}-x_a^{\prime\prime}}{|\vec{x}-\vec{x}^{\prime\prime}|}-\frac{x_a-x_a^{\prime\prime}}{|\vec{x}^{\prime}-\vec{x}^{\prime\prime}|}\right).
\end{equation}
These potentials satisfy the following relations~\cite{Will_1993}:
\begin{equation}
\Delta U=-4\pi\rho,
\end{equation}
\begin{equation}
\Delta \chi = -2 U,
\end{equation}
\begin{equation}
\chi_{,0a}=V_{a}-W_{a},
\end{equation}
\begin{equation}
\Delta V_{a}=-4\pi\rho v_{a},
\end{equation}
\begin{equation}
V_{a,a}=-U_{,0},
\end{equation}
\begin{equation}
\Delta\Phi_1=-4\pi\rho|v|^2,
\end{equation}
\begin{equation}
\Delta\Phi_{2}=-4\pi\rho U,
\end{equation}
\begin{equation}
\Delta\Phi_3=-4\pi\rho\Pi,
\end{equation}
\begin{equation}
\Delta\Phi_4\to-4\pi p,
\end{equation}
\begin{equation}
\chi_{,00}=\mathcal{A}+\mathcal{B}-\Phi_{1},
\end{equation}
\begin{equation}
\Delta\Phi_{W}=4\pi\rho U-4U_{,a}U_{,a}+2U_{,ab}\chi_{,ab}.
\end{equation}

\section{2PN Equations}\label{sec:2PNeq}

The non-vanishing components of the fourth-order equations are $\mathcal{E}_{00}^{(4)}=0$, $\mathcal{E}_{0}^{(4)}=0$ and $\mathcal{E}_{ab}^{(4)}=0$.
With $\ell\equiv\xi b^2$, the first two of these three equations are
\begin{equation}
\begin{aligned}
\mathcal{E}_{00}^{(4)}=&
-\frac{2-\ell}{4}\Delta g_{00}^{(4)}
+\xi b \; \Delta B_0^{(4)}
+\frac{\ell}{4}\left(\Delta g_{aa}^{(4)}-\partial_a\partial_b g_{ab}^{(4)}\right)
\\&
-\frac{1}{2}\kappa \rho \Pi
-\frac{3}{2}\kappa p
-\kappa \rho v_a v_a
-\frac{3\kappa}{8\pi}\partial_0\partial_0 U
\\&
+\frac{\ell \left(-4 \left(3 \ell^2+4\right) \xi ^2+\kappa ^2 (\ell-6) (3 \ell-2)+2 \kappa \xi  (\ell+2) (3 \ell+2) \right)}{4 \pi  (2-\ell)^4}\rho U
\\&
-\frac{4 \kappa\xi\ell \left(5 \ell^2+4 \ell+4\right) -8 \xi ^2\ell \left(3 \ell^2+4\right) +\kappa ^2 \left(\ell^3-24 \ell^2+4 \ell+16\right)}{32 \pi ^2 (2-\ell)^4} \partial_a U \partial_a U
\\&
-\frac{4 \pi \xi b  n_0+\kappa +2 \pi  (\ell-2) m_0}{4 \pi }\partial_0\partial_a V_a
+\frac{4 \pi \xi b  n_0-\kappa +2 \pi  (\ell-2) m_0}{4 \pi }\partial_0\partial_a W_a
\\&
-\frac{\kappa  b^6 (\ell+2)^2 (\kappa -2 \xi )^2 V^{(3)}(0)}{8 \pi ^2 (\ell-2)^4 } U^2=0,
\end{aligned}
\end{equation}

\begin{equation}
\begin{aligned}
\mathcal{E}_{0}^{(4)}=&
\Delta B_0^{(4)}
+\frac{\xi b}{2\kappa}\left(\Delta g_{aa}^{(4)}-\partial_a\partial_b g_{ab}^{(4)}\right)
+\frac{b \xi  (\kappa +\xi )}{\pi  (\ell-2)^2}\rho U
-\frac{b \xi  (\kappa  (3 \ell-14)+8 \ell \xi )}{16 \pi ^2 (\ell-2)^3}\partial_a U \partial_a U
\\&
+\frac{b \kappa  \xi +4 \pi  n_0 (\xi\ell -\kappa  (\ell-2))}{4 \pi  (\kappa  (\ell-2)-\xi \ell  )}\partial_0\partial_a V_a
+\frac{b \kappa  \xi +4 \pi  n_0 (\kappa  (\ell-2)- \xi\ell )}{4 \pi  (\kappa  (\ell-2)- \xi\ell )}\partial_0\partial_a W_a
\\&
-\frac{b^5 (\ell+2)^2  (\kappa -2 \xi )^2 V^{(3)}(0)}{4 \pi ^2 (\ell-2)^4} U^2=0.
\end{aligned}
\end{equation}
For the equation $\mathcal{E}_{ab}^{(4)}=0$, it is more convenient to consider its trace with respect to the 3D background metric, $\delta_{ab}\mathcal{E}_{ab}^{(4)}=0$, which is expressed as
\begin{equation}
\begin{aligned}
\delta_{ab}\mathcal{E}_{ab}^{(4)}=&
\frac{2-\ell}{4}\Delta g_{00}^{(4)}
-\xi b\; \Delta B_{0}^{(4)}
-\frac{4-\ell}{4}\left(\Delta g_{aa}^{(4)}-\partial_a\partial_b g_{ab}^{(4)}\right)
\\&
-\frac{3}{2}\kappa\rho\Pi
+\frac{3}{2}\kappa p
-\kappa \rho v_a v_a
+\frac{3\kappa}{8\pi}\partial_0\partial_0 U
\\&
+\frac{-2 \kappa  \xi  \ell  \left(7 \ell ^2+8 \ell -12\right)+4 \xi ^2 \ell  \left(3 \ell ^2+4\right)+\kappa ^2 \left(9 \ell ^3-28 \ell^2+68 \ell -64\right)}{4 \pi  (\ell -2)^4}\rho U
\\&
+\frac{4 \kappa  \xi  \ell  \left(3 \ell ^2+12 \ell -4\right)-8 \xi ^2 \ell  \left(3 \ell ^2+4\right)+\kappa ^2 \left(-5 \ell ^3+12 \ell^2-68 \ell +64\right)}{32 \pi ^2 (\ell -2)^4}\partial_a U \partial_a U
\\&
+\frac{4 \pi  b \xi  n_0+\kappa +2 \pi  m_0 (\ell -2)}{4 \pi }\partial_0\partial_a V_a
+\frac{-4 \pi  b \xi  n_0+\kappa -2 \pi  m_0 (\ell -2)}{4 \pi }\partial_0\partial_a W_a
\\&
+\frac{3 \kappa   b^6 (\ell +2)^2 (\kappa -2 \xi )^2 V^{(3)}(0)}{8 \pi ^2  (\ell -2)^4}U^2=0.
\end{aligned}
\end{equation}
These three equations can be viewed as a linear system for $\Delta g_{00}^{(4)}$, $\Delta B_{0}^{(4)}$ and $\Delta g_{aa}^{(4)}-\partial_a\partial_b g_{ab}^{(4)}$, which can be solved to yield
\begin{equation}
\begin{aligned}
\Delta g_{00}^{(4)}=&
-\frac{6}{2-\ell}\kappa p
-\frac{2 (\kappa  (\ell +2)-4 \xi  \ell )}{(\ell -2)^2}\rho \Pi
+\frac{8 (\xi  \ell -\kappa )}{(\ell -2)^2}\rho v_a v_a
-\frac{3 \kappa }{2 \pi  (2-\ell)}\partial_0\partial_0 U
\\&
+\frac{\ell  (\kappa -2 \xi ) \left(\kappa  \left(3 \ell ^2+8 \ell +4\right)-4 \xi  \left(3 \ell ^2+4\right)\right)}{\pi  (\ell -2)^5}\rho U
\\&
+\frac{4 \kappa  \xi  \ell  \left(2 \ell ^2+5 \ell -2\right)+8 \xi ^2 \ell  \left(-2 \ell ^2+\ell -2\right)-\left(\kappa ^2 \left(\ell^3+6 \ell ^2+4 \ell -8\right)\right)}{4 \pi ^2 (\ell -2)^5}\partial_a U \partial_a U
\\&
+\frac{\kappa ^2+2 \pi  m_0 (\kappa  (\ell -2)-\xi  \ell )}{\pi  (\kappa  (\ell -2)-\xi  \ell )}\partial_0\partial_a V_a
+\frac{\kappa ^2+2 \pi  m_0 (\xi  \ell -\kappa  (\ell -2))}{\pi  (\kappa  (\ell -2)-\xi  \ell )}\partial_0\partial_a W_a
\\&
-\frac{ b^6 (\ell +2)^3 (\kappa -2 \xi )^3V^{(3)}(0)}{2 \pi ^2  (\ell -2)^6}U^2,
\end{aligned}
\end{equation}

\begin{equation}
\begin{aligned}
\Delta B_{0}^{(4)}=&
\frac{2 \xi b }{2-\ell}\rho \Pi
+\frac{2 \xi b }{2-\ell}\rho v_a v_a
-\frac{\xi b \left(\xi  \left(3 \ell ^2+4\right)-2 \kappa  \left(\ell ^2-\ell +2\right)\right)}{\pi  (\ell -2)^4}\rho U
+\frac{\xi b (\xi  \ell -2 \kappa )}{4 \pi ^2 (\ell -2)^3}\partial_a U \partial_a U
\\&
+\frac{4 \pi  n_0 (\kappa  (\ell -2)-\xi  \ell )-b \kappa  \xi }{4 \pi  (\kappa  (\ell -2)-\xi  \ell )}\partial_0\partial_a V_a
+\frac{4 \pi  n_0 (\xi  \ell -\kappa  (\ell -2))-b \kappa  \xi }{4 \pi  (\kappa  (\ell -2)-\xi  \ell )} \partial_0\partial_a W_a
\\&
-\frac{ b^5 (\ell +2)^3 (\kappa -2 \xi )^2V^{(3)}(0)}{4 \pi ^2  (\ell -2)^5}U^2,
\end{aligned}
\end{equation}

\begin{equation}
\begin{aligned}
\Delta g_{aa}^{(4)}-\partial_a\partial_b g_{ab}^{(4)}=&
-\frac{4 \kappa }{2-\ell}\rho \Pi
-\frac{4 \kappa }{2-\ell}\rho v_a v_a
-\frac{2 \kappa  \left(\kappa  \left(3 \ell ^2-6 \ell +8\right)-2 \xi  \ell  (\ell +2)\right)}{\pi  (\ell -2)^4} \rho U
\\&
+\frac{\kappa  (3 \kappa  (\ell -2)+4 \xi  \ell )}{8 \pi ^2 (\ell -2)^3}\partial_a U \partial_a U
+\frac{\kappa b^6 (\ell +2)^2 (\kappa -2 \xi )^2 V^{(3)}(0)}{\pi ^2 (\ell -2)^5}U^2.
\end{aligned}
\end{equation}

\section{Energy-Momentum Conservation in Vector-Tensor Theories}

Here, we show that in vector-tensor theories, the energy-momentum tensor is automatically conserved when the equation of motion of the vector field is satisfied.
Consider the general action
\begin{equation}
S=S_{EH}
+S_B[g_{\mu\nu},B_\mu]+S_M,
\end{equation}
where $S_{EH}$ is the Einstein-Hilbert action, the action $S_B$ includes the dynamics of the vector field and the coupling between the vector field and the metric, and $S_M$ is the action of matter.
The equation of motion for the vector field is
\begin{equation}
\mathcal{E}^\mu_B = \frac{1}{\sqrt{-g}}\frac{\delta S_B}{\delta B_\mu},
\end{equation}
and the energy-momentum tensor for the vector field is
\begin{equation}
T^{\mu\nu}_B = -\frac{2}{\sqrt{-g}}\frac{\delta S_B}{\delta g_{\mu\nu}}.
\end{equation}
Consider an infinitesimal diffeomorphism generated by the vector field $\xi^\mu$ as $x^\mu\to x^\mu +\xi^\mu$, under which the metric and the vector field transform as
\begin{equation}
\begin{aligned}
\delta g_{\mu\nu}&= \nabla_\mu \xi_\nu + \nabla_\nu \xi_\mu,\\
\delta B_\mu &= \xi^\nu \nabla_\nu B_\mu + B_\nu \nabla_\mu \xi^\nu.
\end{aligned}
\end{equation}
The action $S$ is invariant under diffeomorphism transformations.
Under the infinitesimal diffeomorphism generated by $\xi^\mu$, the variation of $S_{EH}$ vanishes identically by virtue of the Bianchi identities, while the variation of $S_M$ vanishes owing to the conservation of the energy-momentum tensor.
\lh{Indeed, though the bumblebee field might be considered as some kind of matters, we cannot say that the variation of $S_M$ vanishes itself in general. Nevertheless, if the bumblebee sector and ordinary matter sectors decouple, their energy-momentum tensor should be separately conserved, which is exactly the case that we focus on in this work.}
Meanwhile, the variation of the Bumblebee action $S_B$ interrelates the field's energy-momentum tensor with its own equations of motion. Explicitly, it yields
\begin{equation}
\begin{aligned}
\delta S_B =& \int d^4x\sqrt{-g}\Big(
-\frac{1}{2}T_B^{\mu\nu}\delta g_{\mu\nu}+\mathcal{E}_B^\mu \delta B_\mu
\Big)
\\
=& \int d^4x\sqrt{-g}\Big(
-T_B^{\mu\nu}\nabla_\mu \xi_\nu+\mathcal{E}_B^\mu (\xi^\nu \nabla_\nu B_\mu + B_\nu \nabla_\mu \xi^\nu)
\Big)
\\
=& \int d^4x\sqrt{-g}\xi_\nu \Big(
\nabla_\mu T_B^{\mu\nu} + \mathcal{E}_B^\mu \nabla^\nu B_\mu - \nabla_\mu (\mathcal{E}_B^\mu B^\nu)
\Big),
\end{aligned}
\end{equation}
So we have
\begin{equation}
\nabla_\mu T_B^{\mu\nu} = B^\nu \nabla_\mu \mathcal{E}_B^\mu + B^{\mu\nu} \mathcal{E}_{B \mu},
\end{equation}
where $B_{\mu\nu} = \nabla_\mu B_\nu-\nabla_\nu B_\mu$ is the strength tensor.
As we can see, when the equation of motion of the vector field is satisfied ($\mathcal{E}_{B \mu}=0$), the energy-momentum tensor is automatically conserved ($\nabla_\mu T_B^{\mu\nu}=0$).
This conclusion is a natural manifestation of Noether's theorem under diffeomorphism invariance.

\bibliography{refs}

\end{document}